\documentclass[sigconf]{acmart}

\AtBeginDocument{%
  }

\setcopyright{acmlicensed}
\copyrightyear{2018}
\acmYear{2018}
\acmDOI{XXXXXXX.XXXXXXX}

\acmConference[XAI-FIN'24]{Make sure to enter the correct
  conference title from your rights confirmation emai}{November 15th, 2024}{Brooklyn, NY}
\acmISBN{978-1-4503-XXXX-X/18/06}

\usepackage{multirow}
\usepackage{enumitem}
\newlist{myRQs}{enumerate}{1}
\setlist[myRQs, 1]{label=RQ\arabic*}
\usepackage{makecell}

\begin{document}

\title{Differentiable Inductive Logic Programming for Fraud Detection}


\author{Boris Wolfson}
\affiliation{%
  \institution{University of Amsterdam}
  \city{Amsterdam}
  \country{The Netherlands}
}
\email{wolf.brr@gmail.com}

\author{Erman Acar }
\affiliation{%
 \institution{ILLC \& IvI, University of Amsterdam, }
 \city{Amsterdam}
 \country{The Netherlands}}
 \email{e.acar@uva.nl}


\begin{abstract}
In the domain of financial services, fraud detection is one of the tasks that can greatly benefit from explainable AI (XAI) research. Addressing that demand, we investigate the applicability of Differentiable Inductive Logic Programming ($\partial{}$ILP) as an explainable AI approach to Fraud Detection. Although the scalability of $\partial{}$ILP is a well-known issue, we show that with some data curation such as cleaning and adjusting the tabular and numerical data to the expected format of background facts statements,  it becomes much more applicable. While in processing it does not provide any significant advantage on rather more traditional methods such as Decision Trees, or more recent ones like Deep Symbolic Classification,  it still gives comparable results. We showcase its limitations and points to improve, as well as potential use cases where it can be much more useful compared to traditional methods, such as recursive rule learning.  
\end{abstract}






\maketitle



\section{Introduction}
\label{sec:introduction}
Fraudulent activity is as old as money itself \cite{dudek2022symbolic}. 
In modern society, fraud contributes not only to financial loss but also impacts people, industries, entities, services, and the environment. According to the Annual Fraud Report from UK Finance, the overall fraud losses in the UK in 2022 added up to $\pounds$ 1.2  billion, of which fraud losses across payment cards and remote banking had a share of $\pounds$ 726.9 million \cite{fraud-report}. 



Fraudulent action detection classically has been addressed with rule-driven approaches and is still largely a part of the industry practice since explainability in fraud-detection is an important criterion. However, they require laborsome hand crafting when it comes to extending these rules to an ever-changing world with ever-changing business rules, regulations, and even fraud schemes which are dynamic by nature. 

Recently, there has been a rise in the use of machine learning (ML) and data-driven approaches thanks to their high accuracy and adaptivity, however, due to their opaque nature, explaining them requires developing new approaches, giving birth to the field of explainable AI (XAI).  Regarding inherently explainable methods such as Decision Trees (DTs), much of the work done is related to simple rule extraction in the \emph{If-Then} form. This requires a lot of data to train the model, and it is still limited in the generalization and compactness of rules. Generalization is important in the fraud detection area as the available data is usually highly unbalanced towards non-fraudulent transactions. On the other hand, very large rules explaining data do not really help, as they are not easy to progress by the practitioners either. 

A recent research agenda called Neuro-Symbolic (NeSy)  AI focuses on combining the strengths of both worlds; namely developing systems that are both rule-based in nature but also using the strength of flexibility and accuracy of ML approaches \cite{hitzler2022neuro}.





One approach that deals well with small datasets and is known for its ability for generalization is the Inductive Logic Programming (ILP) paradigm, and early and partly less explored rule-based machine learning approach \cite{muggleton1991inductive}. ILP provides a set of rules that explains the input dataset. The dataset consists of positive and negative examples, and the program entails all the positive examples and does not entail any negative ones. Thus, from the machine learning point of view, the ILP system is ultimately a binary classifier that not only explains the data itself but is robust enough to generalize on unseen data. The main disadvantage of ILP, however, is that it does not deal well with noisy, erroneous, or ambiguous data \cite{evans2018learning}. 

More recently, a few works emerged suggesting a neural symbolic extension for implementing the ILP in the very NeSy spirit, one of which is Differentiable Inductive Logic Programming $\partial{}$ILP \cite{evans2018learning}. \citet{evans2018learning} showed that $\partial{}$ILP can provide generalization when applied to noisy data. 

In this paper, we investigate further extension of $\partial$ ILP as a potential application of it to fraud detection, which to the best of our knowledge, is not yet done. In doing so, we look for answers to the following research questions:

\begin{myRQs}[resume]
    \item What is the level of performance concerning other traditional approaches, such as Decision Trees and Deep Symbolic Classification?
    \item What is the trade-off between the size of the rule vs its performance?
    \item To what extent can $\partial$ILP provide explanatory rules by using Recursive Structures for detecting relationships between different agents?
\end{myRQs}

\noindent
To address these questions, first, we develop a synthetic dataset to test the performance of $\partial$ILP, and then we train and evaluate the model on the large simulated dataset called  PaySim \cite{lopez2016paysim}. Then, we compare our approach against existing explainable methods such as Decision Trees and Deep Symbolic Classification~\cite{visbeek2023explainable}. Finally, we look into more complex scenarios such as \textit{Fraud Relationhsip} and \textit{Chain of Fraud} for testing $\partial$ILP's ability to induce recursive rules which is not possible with the classical ML approaches. However, the application of $\partial$ILP is not straightforward as it requires data preprocessing. In order to do so, we introduce a few simple technical adjustments, such as using a database index and binarizing the numerical values.

\section{Related Work}
\label{sec:related_work}

Multiple approaches to derive rules from the tabular data exist - the most famous one is the DT approach, where the $IF-THEN-ELSE$ set of rules is applied to split the data into different categories \cite{breiman1984cart}. Although DTs provide explanatory rules, these rules come in multiple spits,  affecting interpretability. In addition, as was shown by \citet{bengio2010decision} DTs are known to overfit and lack generalization.  

Other approaches use Neural Networks to derive the set of rules, by implementing AND, and OR logical operators by two consecutive layers respectively. The Decision Rules Network (DR-
Net), and the Relational Rule Network (R2N) for example, provide a set of rules in a disjunctive normal form (DNF) as shown in Equation \ref{eq:DNF}  \cite{dr-net, kusters2022differentiable}.

\begin{equation}
    IF \text{ A or B or C ... } THEN \text{ ...}
\label{eq:DNF}
\end{equation}

The work of \citet{collery2022neural} extended this approach by applying an R2N layer as a convolutional window, for discovering patterns for the classification of sequential data. These approaches deliver an inherently explainable set of rules but cannot derive recursive predicates or invent a new predicate. The size of the derived rule formula expresses the importance of recursion. In addition, recursion is useful in generalizing from small datasets. The ability of predicate invention allows algorithms to learn patterns without explicit input from domain experts.




As an alternative to $\partial$ILP, there are other differentiable inductive logic methods, such as differentiable Neural Logic ILP (dNL-ILP) \cite{payani2019inductive} and Meta\textsubscript{Abd} \cite{dai2020abductive}. Both dNL-ILP and $\partial{}$ILP learn target predicates based on the facts defined by the set of predicates and auxiliary predicates. Domain experts usually define auxiliary predicates. The difference between the two approaches lies in how the initial set of rules is generated. $\partial$ILP has several restrictions in so-called language bias, the template that defines how to generate the rules. dNL-ILP has however fewer restrictions and is thus expected to be more scalable. Meta\textsubscript{Abd} approach implements background knowledge as a set of rules, instead of facts and works on images, with no explicit positive and negative examples. All three are capable of implementing recursion and inventing predicates. In this work, we opted to start exploring $\partial$ILP because it is a well-studied and more well-known framework.


On the XAI methods that are applied to fraud detection, there has been a large body of work.  As most well-known industry standards include logistic regression and DTs.  

\citet{hajek2023fraud} compares the performance of the XGBoost (a non-explainable industry standard) framework to other conventional methods, applied on the PaySim dataset, and recent work by \citet{visbeek2023explainable}, introduced a Deep Symbolic Classification (DSC) framework based on symbolic regression, and applied it to fraud detection problem. The DSC learns a set of equational rules, by extending Deep Symbolic regression (DSR)\cite{petersen2019deep}, and their performance results are summarised in Tabel \ref{tab:dsc-xgboost}.

 \begin{table}[h!]
    \centering
    \caption{ DSC and XGBoost performance on PaySim dataset}
    \begin{tabular}{lcc}
\Xhline{2\arrayrulewidth}
 Performance &  DSC\cite{visbeek2023explainable} & XGBoost\\
\Xhline{2\arrayrulewidth}
       Accuracy    & 0.99   & 0.999 \\
       Precision  & \textbf{0.95}   & 0.879 \\
       Recall      & 0.67  &  \textbf{0.806} \\
       F1          & 0.78  &  \textbf{0.841} \\
    \hline

    \end{tabular}
    \label{tab:dsc-xgboost}
\end{table}

Finally, there is an LSTM approach for fraud detection based on the real dataset, which is covered in \cite{alghofaili2020financial}.


\section{The Method: $\partial$ILP}

In introducing the main method, $\partial{}$ILP, we give the intuitive explanation leaving the technical details out and referring the interested reader to \cite{evans2018learning}. 


The rules of an ILP framework are written as clauses of the  following form:
\begin{equation}
    H \leftarrow B_{1}, B_{2}, ...., B_{n},
\end{equation}

where atom $H$ is defined as the head of the clause and the set of atoms $B_{1}, B_{2}, ...., B_{n}$ is defined as the body of the clause. Atom is a predicate applied to a set of terms. Where each term is a variable, for example, a client ID. If all the atoms in the body are true, then the head is necessarily true. The clause is also defined as a definite clause because it has only one head.

For example, the following program defines the set of rules $\mathcal{R}$ for connected relations as the transitive closure of the edge relation:

\begin{equation}
    \begin{aligned}
    connected(X, Y ) &\leftarrow edge(X, Y )\\
    connected(X, Y ) &\leftarrow edge(X, Z), connected(Z, Y).
    \end{aligned}
\end{equation}

Here, the \textit{background knowledge} can be defined as a set of all edges, the \textit{positive examples} as a set of all known connections, and the \textit{negative examples} as a set of examples where the variables are not connected.

$\partial{}$ILP learns a set $\mathcal{R} $  of definite clauses such that the union set of background facts $\mathcal{B} $ and $\mathcal{R} $ entails all the atoms in a set of positive examples $\mathcal{P} $, and does not entail all the atoms in a set of negative examples $\mathcal{N}$:

\begin{equation}
\begin{aligned}
    \mathcal{B}, \mathcal{R} \vDash \gamma, \forall \gamma \in \mathcal{P} \\
    \mathcal{B}, \mathcal{R} \nvDash \gamma, \forall \gamma \in \mathcal{N}
    \end{aligned}
\end{equation}

\subsection{Clause Generations}
Two types of predicates are distinguished in $\partial$ILP: the intensional and extensional predicates. The set of extensional predicates $P_{e}$ are the given predicates from the background knowledge, and the set of intensional $P_{i}$ are the predicates to be learned. The set $P_{i}$ consists of the target predicate and additional auxiliary predicates $P_{a}$.
The central component of $\partial$ILP is based on generating a list of possible definite clauses for intensional predicates, also known as a language bias \cite{tausend1994representing}. The clauses are generated by the so-called rule template $\tau$, which defines the range of clauses to generate. Two clauses define each predicate $p$, therefore there are two templates  $(\tau^1_{p}, \tau^2_{p})$. 

\begin{equation}
    \tau=(n_{\exists_{}},int)
\end{equation}

Where $n_{\exists_{}}$ specifies the number of existentially quantified variables allowed in the clause, and $int$ is a flag that determines whether the atoms in the clause can use intensional predicates $P_{i}$.

The following restrictions are applied when generating clauses:

\begin{itemize}
    \item Each clause consists of exactly two atoms. Where an atom $\gamma$ is any grounded predicate. 
    \item A predicate has a maximal arity of two.
    \item The variable that appears in the head of the clause must appear in its body.
    \item An atom is not used in the same clause's head and body (circular restriction).
    
\end{itemize}

\noindent

It was discovered during development, that the software suite generated rules of the following shape:

\begin{equation}
    \begin{aligned}
    Fraud(X) &\leftarrow Predicate1(X)\\
    Predicate1(X) &\leftarrow Fraud(X), Predicate2(X)
    \end{aligned}
\end{equation}
The Atom $Fraud(X)$ does not appear directly in the body of the same rule but appears after in the body of a dependent Atom $Predicate1(X)$. Therefore, in addition to the formal restrictions mentioned in the work of \citet{evans2018learning}, an extension of the circular restriction was introduced; that is, the target predicate with the same variable can not appear in the head and body of both clauses defining the predicate.

\subsection{Pipeline}
\begin{figure}[h]
\begin{center}
    \includegraphics[width=0.9\columnwidth]{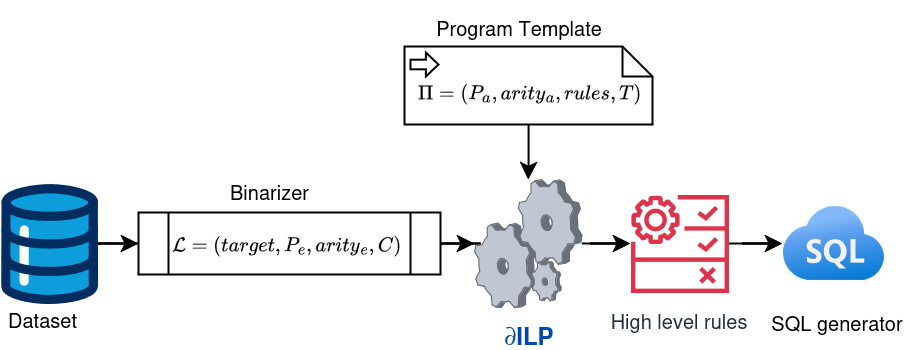}
    \caption{Pipeline $\partial$ILP. The dataset is any tabular dataset. Binarizer converts numerical values to binary values and creates sets of facts $P_e$. Program Template defines a set of clauses to generate. The parameter $rules$ in the program template is a set of rule templates for each intensional predicate including Target. Generated \textit{High-level rules} can be translated to an SQL query afterward}
    \label{fig:dilp-pipeline}
\end{center}
\end{figure}

In this section, we describe different aspects of the implementation. The general overview of the process is described in Figure \ref{fig:dilp-pipeline}.

\subsubsection{Adjustment of $\partial$ILP to use tabular data}
$\partial$ILP is applied to background knowledge consisting of a set of facts, and positive and negative examples of the predicates. The set of facts is binary, therefore the existing input data should be adjusted to the same binary format. A work of \citet{ciravegna2023logicExpNet} about Logic Explained Networks (LEN), suggested discretizing numerical data into different bins to enable a neural learner to use the data. We exploit a similar approach for importing the fraudulent data into the model, by applying a threshold to values to test if it is above or below it. Thus converting a numerical column to a binary one. The binarised column is called a predicate of arity one. Because a predicate is a function of a variable, for the grounding purposes of a variable, it is assigned to an index of the transaction in the dataset, thus explicitly applying the uniqueness of the relevant transaction, for example, $isFraud(X)$, is a predicate $isFraud$ of the transaction $X$.
The DT and the DSC thresholds will be used to binarise the tabular data.

When discussing more complicated predicates with an arity of two, when considering a predicate based on sender and receiver, the grounding of the variable's facts is based on the sender and receiver identity number, assuming that the identity number is unique. For example, $isFraud(X,Y)$, is a predicate $isFraud$ of a transaction between $X$ and $Y$.


\subsubsection{Rule size}
Rule size is defined through the number of possible predicate columns to incorporate in the logical program.

\subsubsection{Program Template}
As discussed before, $\partial$ILP requires an input of Program Template, consisting of the inference step $T$, a set of auxiliary predicates as in Rule size, and rule templates. 

\subsubsection{SQL query generator}
For the 1-arity predicate, an SQL query was generated based on the derived rule and applied to the tabular data. An example rule with a form (Equation \ref{sql}):

\begin{equation}
    \begin{aligned}
        Target(X_0) &\leftarrow Pe_1(X_0), pred1(X_0) \\
        Target(X_0) &\leftarrow Pe_2(X_0), pred2(X_0) \\
        pred1(X_0) &\leftarrow Pe_3(X_0), pred2(X_0) \\
        pred2(X_0) &\leftarrow  Pe_4(X_0). \\
    \end{aligned}
    \label{sql}
\end{equation}

\noindent
Following this rule, the SQL generator gives the query:

\begin{verbatim}
Select 
    Pe4 as pred2,
    Pe3 as pred1,
    Pe2 and pred2 or Pe1 and pred1 as Target,
    from Fraud_Table
\end{verbatim}

\section{Data and the Experimental Setup}
\label{sec:methodology}
A considerable amount of research was done on a simulated scenario PaySim which can be found on Kaggle site\footnote{https://www.kaggle.com/datasets/ealaxi/paysim1}. In this dataset, the fraudulent behavior of the agents aims to profit by taking control of customers' accounts and trying to empty the funds by transferring them to another account and then cashing out of the system. We chose this dataset because it allows us to effectively compare the method to other studies working with the same dataset.

The dataset contains ~6.3 million transactions over one month of simulation, from which the number of fraudulent transactions is 8.2K, giving a ratio of circa 0.13\%. 











\subsubsection{Explorative Data Analysis}
\begin{figure}[h]
\begin{center}
    \includegraphics[width=0.9\columnwidth]{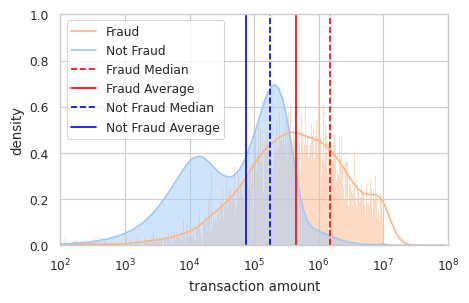}
    \caption{Fraudulent (\textcolor{orange}{Orange}) and Valid (\textcolor{blue}{Blue}) transaction density distributions. Dashed red and blue lines represent medians, and solid lines represent averages of fraudulent and valid transactions}
    \label{fig:fraud-density-dst}
\end{center}
\end{figure}

Figure \ref{fig:fraud-density-dst} shows the fraudulent and valid transaction amount density distributions. Looking into the median and average values, it can be seen that both the Fraudulent and the Valid distributions are shifted with respect to each other. The average and median of fraudulent transactions are higher than the valid ones, meaning the transaction value should play a role in the fraud classification.

\begin{figure}[h]
\begin{center}
    \includegraphics[width=0.9\columnwidth]{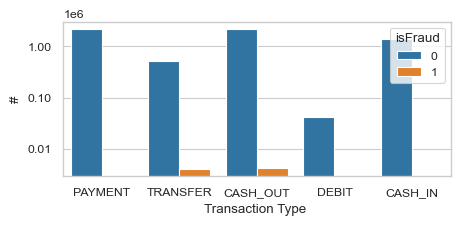}
    \caption{Fraudulent (\textcolor{orange}{Orange}) and Valid (\textcolor{blue}{Blue}) transaction count plot per type of transaction}
    \label{fig:fraud-type-count}
\end{center}
\end{figure}

Additional differences between the transactions are shown in the count plot in Figure 
\ref{fig:fraud-type-count} with a reference to common features in PaySim dataset. Here, it appears that only the features TRANSFER and CASH\_OUT transactions can be fraudulent.

Based on those figures, the expected Rule should be based on the transaction type, as $TRANSFER$ or $CASH\_OUT$ to decide if a transaction is fraudulent, in addition to the transaction amount.
\subsubsection{Feature engeneering}
For the missing value treatment and aggregates calculation we followed the work of \citet{visbeek2023explainable}. 
\paragraph{Missing values treatment}
No $null$ or $NaN$ values were identified in the dataset, however, there are transactions with zero values for old and new balances. Those are transactions from or to the Merchants, who are the Customers with an ID starting with $M$. The balances for these types of transactions were changed to be equal to the amount of a transaction. An additional \textit{external origin}, and \textit{external destination} flags were calculated to mark those types of transactions. In Table \ref{tab:missing_values} there is a schematic overview

\begin{table}[h!]
    \centering
        \caption{Missing values treatment for the transactions with zero balances before or after the transactions}
    \begin{tabular}{ll}
    \Xhline{2\arrayrulewidth}
    Case & Flags \\
    \Xhline{2\arrayrulewidth}
         External origin &  $old\_balance\_origine = amount$ \\
         Old balance origin is zero                & $external\_orig\_flg=True$ \\
    \hline
         External destination &  $new\_balance\_destination = amount$\\
          New balance destination &  $ external\_dest\_flg=True$\\
    \hline

    \end{tabular}

    \label{tab:missing_values}
\end{table}

\paragraph{Test, train, validation}
The transaction data was randomly split into train (85\%) and test(15\%) sets. Subsequently, the training set was split into the validation (15\%) and train (85\%) sets. Assuming the receiver is the fraudulent agent, the splits were performed on groups by the name of the destination. 




\paragraph{Data scaling}
The numerical attributes of the dataset were scaled with a standard scaler fitted on the training dataset.

\paragraph{Aggregates calculation}
We extend the datasets with additional features: 
For each name destination
\begin{itemize}
    \item Average amount of the last 7 and 3 transactions including the current transaction
    \item MAX amount of the last 7 and 3 transactions including the current transaction
\end{itemize}

\subsection{Synthetic generated test data}
\paragraph{Dummy Set}
To understand the role of hyperparameters such as the number of inference steps and program templates, we first test our methods on a small set with dummy synthetic data, the dataset consisted of five binary columns A, B, C, D, and Target. Where the Target column is true if A, B, C, and D are True. The values for A, B, C, and D were randomly generated and are equal to $True$ or $False$. The goal is that the target predicate should be correctly learned with 100\% accuracy.

\paragraph{Additional Fraud scenarios}
The final datasets were created to explore the ability of the method to model recursive predicates. These Fraud scenarios differ from the one described in PaySim, intended to create a rule that can define a chain of fraudulent events. The concept is based on the transitivity case and the graph connection examples from $\partial$ILP. The transitivity case was generated based on the random generation of the transactions between different entities, for the cases where one entity received a fraudulent transaction and then performed a transaction with another entity. This type of transaction scheme is defined as a Chain of Fraud. 

\subsection{Experimental Setup}
The implementation of the $\partial$ILP method is based on the code published by ai-systems \footnote{https://github.com/ai-systems/DILP-Core}. The aggregated features were made with the DuckDb package \footnote{https://duckdb.org/} using SQL queries. The models based on PaySim and Dummy datasets were trained with an X13 Dell laptop, and the recursive examples were trained on the Snellius supercomputer \footnote{https://www.surf.nl/diensten/snellius-de-nationale-supercomputer}.

\subsection{Evaluation metrics}

\subsubsection{Performance}
The performance of the fraud detection framework will be evaluated by the commonly used metrics for classification: accuracy, recall, precision, and F1 score. Additionally, the Matthews correlation coefficient (MCC) score will be evaluated \cite{matthews1975comparison}. MCC takes into account all four values of the confusion matrix, therefore it has a high score only if the classifier can correctly predict the majority of both positive and negative data instances \cite{chicco2020advantages}. It values as: $-1$ all predictions are wrong, $0$ predictions are random, and $+1$ predictions are perfect. $MCC=(tp \cdot tn-fp \cdot fn)/\sqrt{(tp+fp )(tp+fn)(tn+fp )(tn+fn)}$. Due to the imbalanced data, the accuracy score is less relevant, therefore, we focus mainly on precision, F1, and MCC scores.


\section{Results}
\label{sec:results}

The following sections present the results of rule learning over the binarised data. In each case, the number of positive and negative instances is mentioned. The positive instances are where the Target Predicate is $True$, and the negative is where the Target is $False$. The Training set table may contain the rows where all the fact columns are $False$. Those are dropped since $\partial$ILP requires a set of facts that are $True$ over the constant sets. Therefore, the number of instances may be less than the number of data rows.

Derived rules are presented as they appeared from the derivation. In several cases, a head atom is defined by a duplication of an atom in the rule's body. This is explained by the approach's requirement that a body be defined by exactly two atoms.

\subsection{A, B, C, D scenario }
The first experiment was to test the influence of the number of inference steps on $\partial$ILP performance on a dummy dataset. Number of auxiliary predicates: $|p_a|=2$


Table \ref{tab:abcd-rule} summarises the A, B, C, and D experiment results. For the different inference steps, the following rules were achieved

\begin{table}[h!]

    \caption{$\partial$ILP performance on the A, B, C, and D dataset or different inference $T$ steps, a fraction of positive target 7\%, 7 positive and 86 negative examples out of $N=100$ generated rows.}
    
    \centering
    \begin{tabular}{lcccc}
    \Xhline{2\arrayrulewidth}
    Performance & $T=2$   & $T=3$    & $T=5$    & $T=10$ \\
   \Xhline{2\arrayrulewidth}
       Train time sec   & 94 & 156 & 309 & 410 \\  
       Accuracy   & 0.96 & 0.94  & 1 & 1 \\
       Precision  & 0.64 &  0.54 & 1 & 1 \\
       Recall     & 1 & 1 & 1    & 1 \\
       F1         & 0.778 & 0.7  & 1 & 1 \\
       MCC        & 0.78 & 0.71  & 1 & 1 \\
    \hline

    \end{tabular}

    \label{tab:abcd-rule}
\end{table}
$T=2$ inference steps
\begin{equation}
    \begin{aligned}
        Target(X_0) &\leftarrow D(X_0),pred2(X_0) \\
        pred1(X_0) &\leftarrow D(X_0),pred2(X_0) \\
        pred2(X_0) &\leftarrow C(X_0),A(X_0) \\
    \end{aligned}
\end{equation}

Here the rule only partially covers the dataset, because when rephrased it is 
\begin{equation}
    Target(X_0) \leftarrow D(X_0), C(X_0),A(X_0)
\end{equation}
remarkably $pred2$ does not influence the solution

$T=3$ inference steps
\begin{equation}
    \begin{aligned}
        Target(X_0) &\leftarrow pred1(X_0),pred1(X_0) \\
        pred1(X_0) &\leftarrow  pred2(X_0),B(X_0) \\
        pred2(X_0) &\leftarrow  C(X_0),D(X_0) \\
    \end{aligned}
\end{equation}

Here the rule again partially covers the dataset, but with other predicates:
\begin{equation}
    Target(X_0) \leftarrow B(X_0), C(X_0),D(X_0)
\end{equation}

$T=5$ inference steps
\begin{equation}
    \begin{aligned}
        Target(X_0) &\leftarrow pred1(X_0),B(X_0) \\
        pred1(X_0) &\leftarrow  pred2(X_0),A(X_0) \\
        pred2(X_0) &\leftarrow  C(X_0),D(X_0) \\
    \end{aligned}
\end{equation}
which can be rephrased as:
\begin{equation}
    Target(X_0) \leftarrow B(X_0), A(X_0),C(X_0),D(X_0)
\end{equation}

$T=10$ Inference steps. This is an example of a circular dependency between $pred1$ and $Target$ predicates derived during training.
\begin{equation}
    \begin{aligned}
        Target(X_0) &\leftarrow pred1(X_0),B(X_0) \\
        pred1(X_0) &\leftarrow  Target(X_0),Target(X_0) \\
        pred2(X_0) &\leftarrow  A(X_0),D(X_0) \\
    \end{aligned}
\end{equation}
which can be rephrased as:

\begin{equation}
    Target(X_0) \leftarrow B(X_0), Target(X_0),A(X_0),D(X_0)
\end{equation}
 meaning $Target$ depends on $Target$

$T=10$ (Prevent recursion)
\begin{equation}
    \begin{aligned}
        Target(X_0) &\leftarrow pred1(X_0),C(X_0) \\
        pred1(X_0) &\leftarrow  pred2(X_0),A(X_0) \\
        pred2(X_0) &\leftarrow  B(X_0),D(X_0) \\
    \end{aligned}
\end{equation}
which is also a full rule, covering the dataset. But $T=5$ is enough inference steps to cover such a dataset. 

Based on these results, we can already see that in the simulated dataset without errors $\partial$ILP can provide explanatory rules covering the full dataset, achieving ideal performance. Which partially provides an answer to RQ1 concerning a dummy dataset. In addition, the derived rules are not only in if-then form but also relational as opposed to DT, which can support a hierarchical structure. Hence, compared to DTs, $\partial$ILP rules are more expressive and compact, making them more convenient for human experts with a basic logic programming background.

\subsection{PaySim learning}
To provide a more objective answer to RQ1 we trained $\partial$ILP on the PaySim dataset. Training on the full original training set was impossible due to memory limitations. We trained $\partial$ILP on two smaller trainsets, consisting of 50:50\% and 1\% Fraud ratio. 

\subsubsection{Baseline performance}
For establishing a baseline, a DT classifier and a classifier based on the DSC rule were evaluated.
When the DT classifier was tuned to maximize the MCC score. The performances for the training and test sets as summarised in Table \ref{tab:baseline}.
\begin{table}[h!]
    \centering
        \caption{Baseline Performance of DT, and DSC Rule}

    \begin{tabular}{lcccc}
   \Xhline{2\arrayrulewidth}
    \multirow{3}{5em}{Performance} & \multicolumn{2}{c}{DT } & \multicolumn{2}{c}{DSC} \\
    \cline{2-5}
                & training set & test set & training set & test set\\
   \Xhline{2\arrayrulewidth}

       Accuracy   & 0.999  & 0.999  & 0.999   &  0.999 \\
       Precision  & 0.969 & 0.971  & 0.981 &  0.984 \\
       Recall     & 0.647  & 0.665  & 0.493  &  0.501 \\
       F1         & 0.776 & 0.789  & 0.656 &  0.664 \\
       MCC        & 0.792 & 0.803 & 0.695 &  0.702 \\
    \hline

    \end{tabular}
    \label{tab:baseline}
\end{table}

\subsubsection{Parameters}

All the tests were run with $T=5$ inference steps, based on the assumption, that the number of required auxiliary predicates is equal to or is less than two in the "A, B, C, and D" simulated case. 

\subsubsection{DT thresholds}
In Table \ref{tab:decision-tree-50-50} the results for the DT thresholds dataset case are summarized,

\begin{table}[h!]
    \centering
    \caption{$\partial$ILP performance on the dataset induced from DT thresholds. Number of auxiliary predicates $|p_a|=1,2$. The fraction of fraudulent transactions is 50\%. 100 fraudulent and 100 non-fraudulent examples out of $N=200$  rows}
        
    \begin{tabular}{lcccc}
   \Xhline{2\arrayrulewidth}
    \multirow{3}{5em}{Performance} & \multicolumn{2}{c}{$|p_a|=1$} & \multicolumn{2}{c}{$|p_a|=2$} \\
    \cline{2-5}
                & training set & test set & training set & test set\\
   \Xhline{2\arrayrulewidth}

       Train time sec  & 157  & - & 275 &      -      \\
       Accuracy   & 0.540 & 0.999  & 0.540 &  0.999 \\
       Precision  & 1.000 & 1.000  & 1.000 &  1.000 \\
       Recall     & 0.08  & 0.176  & 0.080 &  0.176 \\
       F1         & 0.148 & 0.299  & 0.148 &  0.299 \\
       MCC        & 0.204 & 0.419 & 0.204 &  0.419 \\
    \hline

    \end{tabular}

    \label{tab:decision-tree-50-50}
\end{table}

Derived rules for $|p_a|=1$:
\begin{equation}
    \begin{aligned}
        isFraud(X_0) &\leftarrow external\_dest(X_0),pred1(X_0) \\
        pred1(X_0) &\leftarrow   {amount > 1.297} \\
    \end{aligned}
\end{equation}

Derived rules for $|p_a|=2$:
\begin{equation}
    \begin{aligned}
        isFraud(X_0) &\leftarrow external\_dest(X_0),pred2(X_0) \\
        pred1(X_0) &\leftarrow   pred2(X_0),external\_dest(X_0)\\
        pred2(X_0) &\leftarrow   {amount > 1.297} \\
    \end{aligned}
\end{equation}

both rules express the same rule from the form:
\begin{equation}
    \begin{aligned}
        isFraud(X_0) &\leftarrow external\_dest(X_0),{amount > 1.297} \\
    \end{aligned}
\end{equation}

This led to a high \textit{Precision} score on the test set, meaning there were no False Positives, i.e. \textit{False Fraudulent} transactions, but with high underperforming on the rest of the metrics, resulting in a high number of False Negatives, i.e. a large number of \textit{Fraudulent} transactions were not detected.

\subsubsection{Dataset based on DSC rule}
 Results are presented in Table \ref{tab:symbolic-rule}.

\begin{table}[h!]
    \centering
    \caption{$\partial$ILP performance on the dataset induced from DSC thresholds. Number of auxiliary predicates $|p_a|=1,2$. The fraction of fraudulent transactions is 50\%. 100 fraudulent and 100 non-fraudulent examples out of $N=200$  rows.}
    
        \begin{tabular}{lcccc}
   \Xhline{2\arrayrulewidth}
    \multirow{3}{5em}{Performance} & \multicolumn{2}{c}{$|p_a|=1$} & \multicolumn{2}{c}{$|p_a|=2$} \\
    \cline{2-5}
                & training set & test set & training set & test set\\
   \Xhline{2\arrayrulewidth}

       Train time sec  & 305  & - & 558 &       -     \\
       Accuracy   & 0.715 & 0.9993 & 0.715  &  0.9993 \\
       Precision  & 1     & 0.974  & 1      &  0.974 \\
       Recall     & 0.43  & 0.501  &  0.43  &  0.501 \\
       F1         & 0.601 & 0.662  &  0.601 &  0.662 \\
       MCC        & 0.523 & 0.698  &  0.523 &  0.698 \\
    \hline

    \end{tabular}

    \label{tab:symbolic-rule}
\end{table}



\noindent For both cases the same rule was learned: $isFraud(X_0)$ if \\
$type\_TRANSFER(X_0)$ and $external\_dest(X_0)$ are $True$, explaining the same performance results.

\subsubsection{Training set for 1\% fraction fraud cases}
To estimate the influence of a smaller fraction of the Fraudulent class on the training performance, we extracted 1000 non-fraudulent and 10 fraudulent transactions from DT and DSC thresholds datasets. In Table \ref{tab:1000-10-results} the results for both cases are summarized, the rules were derived for 2 auxiliary predicates

\begin{table}[h!]

    \caption{$\partial$ILP performance on the datasets induced from DSC and DT thresholds. Number of auxiliary predicates $|p_a|=1,2$. The fraction of fraudulent transactions is 1\%. 10 fraudulent and 1000 non-fraudulent examples out of $N=1010$  rows.}
    
    \centering
    \begin{tabular}{lcccc}
   \Xhline{2\arrayrulewidth}
    \multirow{3}{5em}{Performance} & \multicolumn{2}{c}{DT} & \multicolumn{2}{c}{DSC} \\
    \cline{2-5}
                & training set & test set & training set & test set\\
   \Xhline{2\arrayrulewidth}
       Train time sec   & 505 &  - & 859  &  -\\
       Accuracy   & 0.996 & 0.999   & 0.996  & 0.999 \\
       Precision  & 1.000 & 0.974   & 1.000  & 0.973 \\
       Recall     & 0.600 & 0.501   & 0.600  & 0.501 \\
       F1         & 0.750 & 0.662   & 0.75   & 0.662 \\
       MCC        & 0.773 & 0.698   & 0.773  & 0.698 \\
    \hline

    \end{tabular}

    \label{tab:1000-10-results}
\end{table}

Derived rules for DT thresholds dataset:
\begin{equation}
    \begin{aligned}
        isFraud(X_0) &\leftarrow pred1(X_0),pred2(X_0) \\
        pred1(X_0) &\leftarrow pred2(X_0), external\_dest(X_0) \\
        pred2(X_0) &\leftarrow  external\_dest(X_0),type\_TRANSFER(X_0) \\
    \end{aligned}
\end{equation}

Derived rules for DSC thresholds dataset:
\begin{equation}
    \begin{aligned}
        isFraud(X_0) &\leftarrow type\_TRANSFER(X_0),pred1(X_0) \\
        pred1(X_0) &\leftarrow  pred2(X_0),external\_dest(X_0) \\
        pred2(X_0) &\leftarrow  type\_TRANSFER(X_0),type\_TRANSFER(X_0) \\
    \end{aligned}
\end{equation}

Remarkably, the rule derived on the DT dataset is the same as for the DSC dataset one and it boils down to $  isFraud(X_0) \leftarrow  external\_dest(X_0),type\_TRANSFER(X_0) $, therefore the same performance is observed between two approaches.

\subsubsection{Dataset based on DT thresholds including the negation}
When deriving binary predicates based on the DT thresholds, the negated branches are not being checked by $\partial$ILP, as it cannot generate the negated predicates. Therefore additional predicates were generated, equaled to the negation of the first set. The Fraud templates should support the DT paradigm, which defines rules as \textit{if A then B, else if not A then C}. The test was executed with two auxiliary predicates set up on both the 50\% and the 1\% fraud dataset.

\begin{table}[h!]
    \centering

   \caption{$\partial$ILP performance on the dataset induced from DT thresholds including negation. Number of auxiliary predicates $|p_a|=2$. The fraction of fraudulent transactions is 1\%. 10 fraudulent and 1000 non-fraudulent examples out of $N=1010$  rows.}

    \begin{tabular}{lcccc}
   \Xhline{2\arrayrulewidth}
    \multirow{3}{5em}{Performance} & \multicolumn{2}{c}{50 \%} & \multicolumn{2}{c}{1 \%} \\
    \cline{2-5}
                & training set & test set & training set & test set\\
   \Xhline{2\arrayrulewidth}
      Train time sec   & 1181 &  - & 1958  &  -\\
       Accuracy   & 0.715 & 0.999   & 0.996  & 0.999 \\
       Precision  & 1.000 & 0.974   & 1.000  & 0.973 \\
       Recall     & 0.430 & 0.501   & 0.600  & 0.501 \\
       F1         & 0.601 & 0.662   & 0.75   & 0.662 \\
       MCC        & 0.523 & 0.698   & 0.773  & 0.698 \\
    \hline

    \end{tabular}

    \label{tab:decision-tree}
\end{table}

Derived rules for 50\%:
\begin{equation}
    \begin{aligned}
        isFraud(X_0) &\leftarrow NOT\{oldbalanceDest > -0.007\}(X_0),pred2(X_0) \\
        isFraud(X_0) &\leftarrow pred2(X_0),{amount > 1.297}(X_0) \\
        pred1(X_0) &\leftarrow NOT\{oldbalanceDest > -0.007\}(X_0),pred2(X_0) \\
        pred2(X_0) &\leftarrow  external\_dest(X_0),type\_TRANSFER(X_0) \\
    \end{aligned}
\end{equation}

Derived rules for 1\%:
\begin{equation}
    \begin{aligned}
        isFraud(X_0) &\leftarrow type\_TRANSFER(X_0),pred1(X_0) \\
        isFraud(X_0) &\leftarrow pred1,pred2(X_0) \\
        pred1(X_0) &\leftarrow  external\_dest(X_0),pred2(X_0) \\
        pred2(X_0) &\leftarrow  external\_dest(X_0),type\_TRANSFER(X_0) \\
    \end{aligned}
\end{equation}

Also in this case the performance is similar to the DSC thresholds-based case, in a deeper analysis, only the 
\begin{equation*}
    isFraud(X_0) \leftarrow  external\_dest(X_0),type\_TRANSFER(X_0)
\end{equation*}
played a role, meaning the other rules were not engaged or also were $True$ when the main condition was $True$.

\subsubsection{Applying a large number of predicates}  
\label{many-predicates}
In order to provide an answer for RQ2 we tested various numbers of auxiliary predicates from two to eight to cover all possible columns in the DT dataset. The results showed the same performance as reported for two predicates. Meaning it did not show a difference in performance.

\subsection{Learning the Recursive Structures}
In the final set of experiments aimed at answering RQ3, we tested the ability of $\partial$ILP to model recursive predicates. In this scenario, the knowledge of fraud is already known, for example by the superior classification algorithm, but there is an intention to derive patterns in the data, as in the case of money laundering patterns.

\subsubsection{Fraud Relationship}
In this scenario, the intention was to derive based on the dataset a rule, that can find a fraudulent relationship, based on background knowledge. The tabular dataset was prepared based on the example of Graph Connectedness from the original paper.
All the transactions are fraudulent.

The derivation of a rule took 957 seconds, based on the input consisting of 4 Facts and 9 Positive examples:

\begin{equation}
    \begin{aligned}
        Fraudsters(X, Y) &\leftarrow Fraud(X, Y) \\
        Fraudsters(X, Y) &\leftarrow Fraud(Z, Y), \\
        & Fraudsters(Z, X) \\
    \end{aligned}
\end{equation}

Here the first Rule is $Fraudsters(X, Y) \leftarrow Fraud(X, Y)$ is inherent based on the dataset, as for the transaction between 1 and 2 ($Fraud(1,2)$). The second rule is reflected by the example of Fraud transactions between 1 and 2 ($Fraud(1,2)$) and 1 and 4 are Fraudsters from the facts ($Fraudsters(1,4)$), therefore it explains 2 and 4 are Fraudsters ($Fraudsters(2,4)$). Although it does appear in the facts, here we can see that it can be generalized to the examples which do not appear in the training set. 



\subsubsection{Chain of Fraud}
This is an extension of the previous example, with a rationale to create a dataset with a chain of events, to find a rule for the transaction between three parties, participated in fraud.

The derivation of a rule took 1355 seconds, based on the input consisting of 36 facts, 5 positive and 21 negative examples:

\begin{equation}
    \begin{aligned}
        Fraud\_Chain(X, Y) &\leftarrow Fraud(Z, X), Transaction(X, Y) \\
    \end{aligned}
\end{equation}

That rule can be translated as \textit{ "There is a chain of Fraud between $X$ and $Y$ if there is a transaction from $X$ to $Y$ and a fraud event between any $Z$ and $X$. }

\begin{figure}[h]
\begin{center}
    \includegraphics[width=0.8\columnwidth]{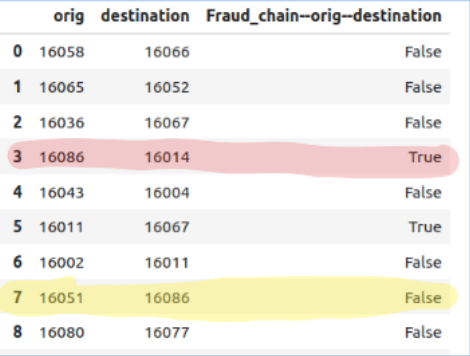}
    \caption{Fraud chain example: The transaction  from agent 16051 to 16086, highlighted in yellow, is not a fraud chain, while the one in red, from 16086 to agent 16014, is.}
    \label{fig:chain-example}
\end{center}
\end{figure}

There is a fraudulent transaction \#8 from customer 16051 to customer 16086 and also a regular transaction \#4 from customer 16086 to customer 16014, which satisfies the $Fraud(16051,16086)$ and $Transaction(16086, 16014)$, and hence satisfies  the head of the rule: 
$Fraud\_Chain(16086, 16014)$, see Figure \ref{fig:chain-example}.

\section{Discussion}
\label{sec:discussion}
In this section, we interpret the obtain results from the angles of performance, scalability, circular dependency and interpretability in general. 
\subsection{Performance}
\subsubsection{Dummy dataset}
When learning the rules from an error-free dataset $\partial$ILP succeeds in finding the logical rules. The number of inference steps can be seen as a hyperparameter for tuning as in the case of $A,B,C,D$ rule learning.

\subsubsection{PaySim dataset}
Regardless of the way of training on the DT or DSC converted Datasets, $\partial$ILP was in line with the performance on the test set in terms of all the metrics compared to the performance of the DSC approach. The DT Classifier performed about 10\% better than $\partial$ILP achieved in terms of F1, Recall, and MCC metrics, as shown in Table \ref{tab:comparison-to-baseline}, but $\partial$ILP provided a more compact rule. The reported Recall and F1 of DSC in work of \citet{visbeek2023explainable} is better than the performance achieved by applying the same rule on the dataset, that could be due to the different dataset splits. 

 \begin{table}[h!]
    \centering
    \caption{Performance comparison between DT, DSC and $\partial$ILP}
    \begin{tabular}{lccccc}
    \Xhline{2\arrayrulewidth}
      Performance &  $\partial$ILP  &   DT   & DSC & DSC\cite{visbeek2023explainable} & XGBoost\\
      \Xhline{2\arrayrulewidth}
       Accuracy   & 0.999  & 0.999          &  0.999  & 0.99   & 0.999 \\
       Precision  & 0.973  & 0.971          & \textbf{0.984} & 0.95   & 0.879 \\
       Recall     &  0.501 & 0.665          & 0.501 & 0.67  &  \textbf{0.806} \\
       F1         & 0.662  & 0.789          & 0.664  & 0.78  &  \textbf{0.841} \\
       MCC        & 0.698  & \textbf{0.803} & 0.702 & -    & - \\

    \end{tabular}
    \label{tab:comparison-to-baseline}
\end{table}

\subsubsection{Data Conversion} We covered two approaches to utilizing tabular data for ILP methods, based on transaction IDs, or based on agent IDs. We showed how to prepare binary data based on the DT  or DSC thresholds, and what Program Templates to use.
When $\partial$ILP was applied to the dataset prepared by DT thresholds, the number of auxiliary predicates, and a fraction of Fraud influenced the performance. We saw that for data including only positive (non-negated) columns and 50-50\% split, the performance based on the Recall, F1, and MCC were lower than for the rest of the approaches (0.176, 0.299. 0.419) vs (0.501, 0.662. 0.698), see Table \ref{tab:decision-tree-50-50} and \ref{tab:decision-tree}.
Therefore it is safe to claim that the general approach to converting a dataset is by including both negated and non-negated columns in terms of above and below a specific threshold and defining a template for a target predicate rule that can combine both. In this way, the created dataset covers both possibilities. Finally, the dataset has to reflect a real ratio of the Target predicate in the original dataset.

\subsubsection{Recursion and Connectivity}
 $\partial$ILP succeeded in learning the rule for the Fraud Relationship and the Chain of Fraud cases. 

\subsection{Scalability}
The work of\citet{evans2018learning} addressed the reason for memory consumption, it depends on the size of invented predicates and constants sets. The constant set size is proportional to the dataset length. In addition, memory size also depends on the amount of the generated clauses. In the case where templates define the predicates with an arity of two, $\partial$ILP generates a very large dataset of clauses. This is the cause for the large training time when applied to relatively small datasets as in the case of connectivity and recursion.

\subsection{Circular Dependency}

We showed that $\partial$ILP can induce rules that have circular dependencies between predicates, the chance is higher in the case of one arity predicates. We implemented the removal of circular dependency on the target predicate. Yet we suspect that because there is no restriction to using two head atoms from two different clauses in each other bodies, there still is a chance of circularity in other predicates. Implementing additional circular restrictions on the predicate generation could be another future topic to investigate.

\subsection{Interpretability}
The generated set of rules are in the form of implications (\emph{If-Then} sentences), and have higher expressivity, as they are relational and can even express circular dependencies, as opposed to  DT rules of a rather flat hierarchy with many branches. Hence, we assume they are readable by a human with a basic logic programming background (still to be confirmed by future research).

\section{Conclusion}
\label{sec:conclusion}

In this article, we have investigated the application of $\partial$ILP on fraud detection. Currently, the method seems insufficient for application in real-world scenarios; a key limitation is the usage of binary predicates, requiring the creation of the binary features from the numerical data with the help of other classification techniques. An additional overhead comes from the high complexity and memory consumption required to create predicates for the approach.

Regarding RQ1, we have seen that training on a small part of the PaySim dataset, $\partial$ILP cannot outperform the techniques used for data transformation to the Boolean format, required by $\partial$ILP. However, $\partial$ILP showed the ability to provide a more explainable rule than the DT, when examining the size of the formula. Training of the full PaySim dataset was not possible due to the memory limitation, therefore it is not clear what would be the performance when trained on a larger training set (hence part of future research). 

Concerning RQ2,  we have seen that this parameter can play the role of underfitting in the worst-case scenario as a rule will not cover all the relationships in the data. In addition, as shown in the case when applying a varying number of rules, the performance stayed the same when learning the DT dataset (\ref{many-predicates}).

Finally, with regards to RQ3, for more complicated scenarios that require building recursive rules, $\partial$ILP successfully provided explanatory rules to the dataset that the methods such as DT, or DSC by the definitions of the approaches, were not able to derive. 

\bibliographystyle{ACM-Reference-Format}
\bibliography{references}

\end{document}